\documentclass[reprint,amsmath,amssymb,aps,prb,showpacs]{revtex4-1}
\usepackage{amsmath}
\usepackage{color}
\usepackage{amssymb}
\usepackage{amsfonts}
\usepackage{txfonts}
\usepackage{graphicx}
\usepackage{bm}
\usepackage{hyperref}
\begin{document}
\title{Obtaining localization properties efficiently using the Kubo-Greenwood formalism}
\author{Andreas Uppstu}
\author{Zheyong Fan}
\email[Corresponding author: ]{zheyong.fan@aalto.fi}
\author{Ari Harju}
\affiliation{COMP Centre of Excellence and Helsinki Institute of Physics,
Department of Applied Physics, Aalto University, Helsinki, Finland}
\date{\today}

\begin{abstract}
We establish, through numerical calculations and comparisons with a recursive Green's function based implementation of the Landauer-B\"uttiker formalism, an efficient method for studying Anderson localization in quasi-one-dimensional and two-dimensional systems using the Kubo-Greenwood formalism. Although the recursive Green's function method can be used to obtain the localization length of a mesoscopic conductor, it is numerically very expensive for systems that contain a large number of atoms transverse to the transport direction. On the other hand, linear-scaling has been achieved with the Kubo-Greenwood method, enabling the study of effectively two-dimensional systems. While the propagating length of the charge carriers will eventually saturate to a finite value in the localized regime, the conductances given by the Kubo-Greenwood method and the recursive Green's function method agree before the saturation. The converged value of the propagating length is found to be directly proportional to the 
localization length obtained from the exponential decay of the conductance.
\end{abstract}

\pacs{72.80.Vp, 72.15.Rn, 73.23.-b, 05.60.Gg}
\maketitle

\section{Introduction}
Localization of waves is an intriguing physical phenomenon that can be encountered in many fields of physics. Anderson localization \cite{Anderson1958} of the charge carriers in mesoscopic conductors leads to an exponentially decreasing conductance with sample length, a phenomenon that may be encountered in sufficiently phase-coherent nanomaterials, such as graphene or carbon nanotubes. In graphene, the phase-coherence length can reach several microns at low temperatures,\cite{Berger2006} and due to the large length scales involved, computational simulation of localization is often challenging.

There are two main numerical methods for studying electronic transport at the quantum level, namely, the  Landauer-B\"uttiker formalism \cite{Datta1995} and Kubo-Greenwood (KG) method.\cite{Kubo1957, Greenwood1958} In the Landauer-B\"uttiker formalism, one studies a system connected to two or multiple terminals, and relates the conductance of the system to the probabilities of the charge carriers to transmit from one terminal to another. The transmission probabilities can be evaluated using Green's function techniques, with the recursive Green's function (RGF) technique being the numerically most efficient implementation for a two-terminal system. However, although the computational effort scales linearly with the length of the system, it scales cubically with the width of the system. On the other hand, in the tight-binding approximation, the (real-space) KG method \cite{Mayou1988, Mayou1995, Roche1997, Triozon2002} can be implemented so that the computational effort scales linearly with the total number of atoms in the system, which enables the study of much wider systems. Although it cannot take into account effects that occur due to contacts or contact-device interfaces, the KG method is in principle very suitable for obtaining intrinsic diffusive transport properties of the device, such as conductivity and mean free path. 

Both the RGF and KG methods have been widely used to study electronic transport in disordered graphene in the diffusive transport regime, while most direct studies of the localized regime have utilized the RGF formalism. \cite{Avriller2006,Li2008,Evaldsson2008,Mucciolo2009,Zhang2009,Bang2010,Uppstu2012,Lee2013} The main difficulty of using the KG method in the localized regime is a lack of a length scale, which is crucial for characterizing the scaling behavior of conductance. This makes the original KG formula not very suitable for studying mesoscopic transport, while a `mesoscopic KG formula' \cite{Fisher1981, Economou1981, Baranger1989, Nikolic2001} was found to be equivalent to the RGF formalism. However, it has been realized recently that a definition of length is possible by recasting the KG formula into a time-dependent Einstein formula,\cite{Roche1997, Triozon2002, Markussen2006, Leconte2011, Radchenko2012, Lherbier2012,  Cresti2013, Fan2013} in which the conductivity is expressed as a time-derivative of the mean square displacement (MSD). This enables one to define a length in terms of the square root of the MSD, which has a clear interpretation in the ballistic regime. The possibility of defining a length combined with the linear-scaling techniques developed in the KG formalism, which still lack successful counterparts in the RGF formalism, makes the KG method very promising for simulating electronic transport in realistically-sized graphene-based systems. However, the relationship between the KG and the RGF methods in the localized regime has not been exactly determined. So far, it is known that the MSD given by the KG method will saturate to a finite value, thought to equal the localization length.\cite{Triozon2000, Takada2013} On the other hand, before this saturation there is a regime where the conductance given by the KG method decays exponentially, roughly equaling the conductance given by the RGF method.\cite{Fan2013}

In this paper, we present methods to obtain the localization properties of a mesoscopic conductor using the KG method, using a hexagonal graphene lattice as a testbed. In the localized transport regime, the conductance decreases exponentially with the length of the conductor, with the exponential decrease being quantified by the localization length $\xi$. Thus, one may perform an exponential fit to the simulated conductance to obtain $\xi$, which is indeed a standard way to find $\xi$ using the RGF method. Thus, we first compare the conductance given by the KG method with the one acquired using the RGF method, and establish how an accurate correspondence between the results from these two methods can be achieved, before the saturation of the MSD. However, the selection of the fitting region to obtain $\xi$ is not completely unambiguous when the KG method is used. We find a direct relation between the converged value of the MSD and the localization length, which provides a numerically more efficient and accurate method to obtain $\xi$.  Using a graphics processing unit (GPU) \cite{Harju2013} to perform the KG simulations, we are able to reach the localized regime and the saturation of the MSD without especially time-consuming computations.

\section{Models and Methods}

As a model system, we have chosen graphene nanoribbons (GNRs) with vacancy-type defects, modeled by randomly removing carbon atoms according to a prescribed defect concentration, which is defined to be the number of vacancies divided by the number of carbon atoms in the pristine system. Our calculations are performed in both armchair and zigzag edged graphene nanoribbons (AGNRs and ZGNRs, respectively), both in the quasi-1D region, where the width $W \ll \xi$ and in the effectively 2D region, where $W \gg \xi$, as well as in the transitional region. An AGNR with $N_x$ dimer lines along the zigzag edge and a ZGNR with $N_y$ zigzag-shaped chains across the armchair edge are abbreviated as $N_x$-AGNR and $N_y$-ZGNR, respectively. The carbon-carbon distance is chosen to be 0.142 nm. We use the nearest-neighbor tight-binding approximation to obtain the Hamiltonian $H$ of the system, which gives in the notion of second quantization
\begin{equation}
H=-t_0 \sum_{\langle i,j \rangle} a^{\dagger}_{i}a_{j},
\end{equation}
where the hopping parameter $t_0$ is set to $2.7$ eV and the sum runs over all pairs of nearest neighbors. Spin degeneracy is assumed.

In the time-dependent Einstein formula \cite{Mayou1988, Mayou1995, Roche1997, Triozon2002, Markussen2006, Leconte2011, Radchenko2012, Lherbier2012, Fan2013} derived from the KG formula, the zero-temperature electrical conductivity at Fermi energy $E$ and correlation time $t$ may be obtained as a time-derivative of the MSD $\Delta X^2(E, t)$, i.e.,
\begin{equation}
\label{equation:REC}
\sigma(E, t) = e^2 \rho(E) \frac{d}{2 d t}
            \Delta X^2(E, t),
\end{equation}
where
\begin{equation}
\Delta X^2(E, t) =
\frac{\textmd{Tr}\left[\delta(E-H)\left(X(t) - X\right)^2\right]}
{\textmd{Tr}\left[\delta(E-H)\right]},
\label{MSD}
\end{equation}
$X(t) = \exp(i H t/\hbar) X \exp(-i H t/\hbar)$ being the position operator $X$ in the Heisenberg representation,
and
\begin{equation}
\rho(E) = \textmd{Tr}\left[\frac{2 }{\Omega}\delta(E-H)\right]
\label{DOS}
\end{equation}
is the density of states (DOS) with spin degeneracy included. The factor of 2 in the denominator of Eq.~(\ref{equation:REC}) should be there to make this equation consistent with the original KG formula. Also, except for pure diffusive transport, the time-derivative in this equation cannot be substituted by a time-division.  For a three-dimensional system, $\Omega$ is the volume of the system. However, for both quasi-1D GNRs and truly 2D graphene, we can simply omit the dimension perpendicular to the basal plane and take $\Omega$ as the area of graphene sheet. As a result, the conductivity has the same dimensionality as the conductance. We further define
\begin{equation}
\lambda(E,t) \equiv \rho(E) \Delta X^2(E, t) = \textmd{Tr}\left[ \frac{2}{\Omega}\delta(E-H) \left(X(t) - X\right)^2\right]
\end{equation}
to facilitate our discussion. The details of the numerical techniques used in the KG simulations have been presented in Ref. \onlinecite{Fan2013}.

One may note that the quantity calculated by Eq. (\ref{equation:REC}) is the conductivity. However, the proper quantity characterizing mesoscopic transport is the conductance. Fortunately, there exists a definition of length in terms of the MSD,
\begin{equation}
 \label{equation:length_definition}
 L(E, t) = 2 \sqrt{\Delta X^2(E, t)},
\end{equation}
which enables converting conductivity to conductance: 
\begin{equation}
 \label{equation:g}
 g(E) = \frac{W}{L(E, t)} \sigma(E,t).
\end{equation}
The conductance defined in this way corresponds exactly to the two-terminal conductance for a ballistic conductor, apart from some numerical problems around the van Hove singularities \cite{Markussen2006}. In general, $g(E)$ is length-dependent and it is one of our purposes to find out how accurate correspondence between independent RGF calculations can be achieved by using the above two definitions given by Eqs.~(\ref{equation:length_definition}) and (\ref{equation:g}). The factor of 2 in Eq.~(\ref{equation:length_definition}) has been justified in Ref. \onlinecite{Fan2013}. Intuitively, it means that $\sqrt{\Delta X^2(E, t)}$ equals the diffusion length only in one transport direction, and the factor of 2 accounts for the opposite direction. We further note that in the purely ballistic regime, this definition of length leads to a length-independent conductance can be derived as $g(E) = e^2 \rho(E) v(E) W/2$, with $v(E)$ being the Fermi velocity, which is equivalent to the textbook formula.\cite{Datta2012} The factor of 2 in this textbook formula has a similar meaning: only half of the carriers at the Fermi energy move along transport direction. Remarkably, the conductance calculated by Eq.~(\ref{equation:g}) does give the correct results for quasi-1D ballistic conductors, except for some numerical inaccuracies around the van Hove singularities. \cite{Markussen2006,Fan2013} We stress that the length defined in Eq. (\ref{equation:length_definition}) is the propagating length of electrons, rather than the length of the cell used in a specific simulation. The simulation cell only needs to be sufficiently large to eliminate finite-size effects resulted from the discreteness of the  spectrum of a finite system.

In the RGF method,\cite{Datta1995} the conductance is calculated using the Green's function, defined as
\begin{equation}
G(E)=\left[EI-H-\Sigma_L(E_L)-\Sigma_R(E_R)\right]^{-1},
\end{equation}
where $I$ is the unit matrix and $\Sigma_{L/R}(E_{L/R})$ is the self-energy of the corresponding lead. The Fermi energy $E_{L/R}$ of the leads can be set to the same value as in the device, $E$, or to an arbitrary value. In this work, we set it either to $E$ or to a value of 1.5 eV, where the latter value is used to simulate metallic contacts with a high number of propagating modes. The self energy matrices may be obtained through different methods, e.g., using an iterative scheme.\cite{Guinea1983} The matrices $\Gamma_{L/R}$, that describe the coupling to the leads, are obtained from
\begin{equation}
\Gamma_{L/R}=i\left[ \Sigma_{L/R}-\Sigma_{L/R}^\dagger \right],
\end{equation}
and the conductance is then computed using
\begin{equation}
g(E)=g_0\text{Tr} \left[ \Gamma_R G(E) \Gamma_L G^\dagger(E) \right],
\end{equation}
where $g_0\equiv 2e^2/h$ is the quantum of conductance.

The conductance of a mesoscopic system depends on the exact configuration of disorder. This effect is especially significant in the localized regime, where variations in the disorder configuration give rise to conductance fluctuations of several orders of magnitude. Thus the conductance and the DOS are not uniquely determined by the defect concentration, and methods for ensemble averaging are required. In the diffusive regime, the distribution of the conductance values is roughly normal, and arithmetic averaging is the preferred method. On the other hand, in the localized regime the conductance values tend toward a log-normal distribution. Thus, in the localized regime the typical value of the distribution, defined as $g_\text{typ}(E) \equiv \exp \langle \ln g(E) \rangle$, describes the ensemble in a statistically meaningful way.\cite{Anderson1980}

In the RGF formalism the average and typical conductances are straightforward to obtain, but one has to note that in  the localized regime a much larger ensemble is required for obtaining an accurate value of the average conductance. On the other hand, when computing the conductance using the KG method, the trace operations of Eqs. (\ref{MSD}) and (\ref{DOS}) correspond to arithmetic averaging. In order to be able to obtain typical values for the conductance and DOS using the KG formalism, we have defined
\begin{equation}
\lambda_i(E,t)   \equiv \langle i \mid \frac{2}{\Omega}\delta(E-H) \left(X(t) - X\right)^2 \mid i \rangle,
\end{equation}
and 
\begin{equation}
\rho_i(E) \equiv \langle i \mid       \frac{2 }{\Omega}\delta(E-H)           \mid i \rangle,
\end{equation}
where $| i \rangle $ is a wave function located at site $i$. Thus by varying $i$ and the defect configuration, we obtain an ensemble of values for $\lambda(E,t)$ and $\rho(E)$. The ensemble is represented by the average values $\lambda_\text{ave} \equiv \langle \lambda \rangle $ and $\rho_\text{ave} \equiv \langle \rho \rangle$, as well as by the typical values $\lambda_\text{typ} \equiv \exp \langle \ln \lambda  \rangle$ and $\rho_\text{typ} \equiv \exp \langle \ln \rho  \rangle$. Thus when utilizing the KG method, we obtain the average and typical conductances from
\begin{equation}
\label{REC_ens}
g_\text{ave/typ}(E, t) = \frac{W}{L_\text{ave/typ}(E, t)} e^2 \frac{d}{2 d t}
            \lambda_\text{ave/typ}(E, t),
\end{equation}
where the length $L_\text{ave/typ}(E,t)$ is obtained from the expression
\begin{equation}
\label{length}
L_\text{ave/typ}(E,t)=2\sqrt{\Delta X^2_\text{ave/typ}(E, t)},
\end{equation}
where $\Delta X^2_\text{ave/typ}(E, t) = \lambda_\text{ave/typ}(E, t) / \rho_\text{ave/typ}(E) $.

\section{Direct comparison of the conductance in the localized regime}

\begin{figure}
\begin{center}
  \includegraphics[width=\columnwidth]{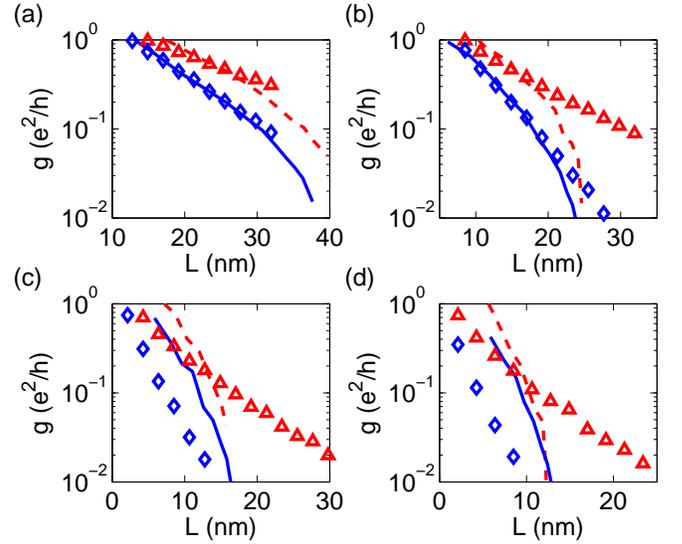}
  \caption{(Color online) Average and typical conductances of 65-GNR at 0.4 eV (a), 0.3 eV (b), 0.2 eV (c) and 0.1 eV (d). The lines correspond to KG results and the symbols to RGF results. The triangles and dashed lines indicate the average conductance, while the diamonds and solid lines indicate the typical conductance. The defect concentration is $1 \%$ and the ensemble size is 10000.}
  \label{G_comp}
\end{center}
\end{figure}

\begin{figure}
\begin{center}
 \includegraphics[width=\columnwidth]{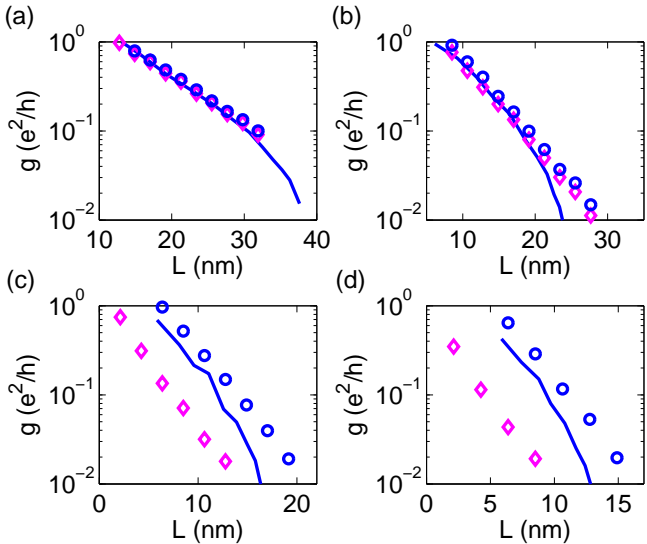}
  \caption{(Color online) Typical conductances of 65-GNR at 0.4 eV (a), 0.3 eV (b), 0.2 eV (c) and 0.1 eV (d). The solid lines indicate KG results, while the symbols indicate RGF results. The circles have been computed by setting the Fermi energy of the semi-infinite leads to 1.5 eV, while the diamonds have been computed using the same Fermi energy in the leads as in the device. The defect concentration is $1 \%$ and the ensemble size is 10000.}
  \label{eva}
\end{center}
\end{figure}

When the length of the system is much larger than $\xi$, the conductance decays exponentially. Early work by Landauer \cite{Landauer1970} showed that $g_\text{ave}$ decays as $\sim \exp (-L/2\xi)$, whereas Anderson \textit{et al.} \cite{Anderson1980} noted that $g_\text{typ}$ decays faster, i.e. as $\sim \exp (-L/\xi)$. Let us first study the conductance far away from the charge neutrality point (CNP), with the number of propagating modes being large in this region. We compute the average and typical conductances using both the RGF and KG methods, obtaining the conductance from Eq. (\ref{REC_ens}) in the latter case. In both methods, we have obtained the values from ensembles of 10000 random defect configurations, and the simulation cell is set to be about 200 nm long in the KG simulations. The vacancy concentration is set to 1\%. Figures \ref{G_comp}(a) and (b) show a direct comparison between results from the RGF and KG methods in the localized regime, for a 65-AGNR at $E=0.4$ eV and $E=0.3$ eV, 
respectively. As the MSD finally reaches an upper bound, the conductivity given by Eq.~(\ref{equation:REC}) drops eventually to zero super-exponentially.\cite{Fan2013} Before the super-exponential decay of the conductance, both the typical and the average conductances given by the KG method decrease exponentially and in good agreement with the corresponding values given by the RGF method. From the slopes of the typical conductances, the localization lengths can be obtained as roughly 8 nm at $E=0.4$ eV and 5 nm at $E=0.3$ eV. Thus it can be seen that the regime where $g_\text{typ}$, given by the KG method, decays exponentially extends roughly to four times $\xi(E)$, with $g_\text{ave}$ exhibiting a somewhat shorter regime of exponential decay.

At energies closer to the CNP, a clear discrepancy between results from the two methods rises, as shown in Figs. \ref{G_comp} (c) and (d). This may be explained by considering the number of transmission channels, which in pristine GNRs is low close to the CNP. Thus the semi-infinite leads limit the conductance of the complete system, as the disordered device exhibits impurity induced states \cite{Pereira2006,Pereira2008,Robinson2008,Wehling2010,Cresti2013} which may enhance the conductance. The limiting effect of the leads is reduced by considering highly doped leads instead,\cite{Katsnelson2006, Tworzydlo2006,Cresti2013} which allows for a higher conductance of the complete system. This is illustrated in Fig. \ref{eva}, which compares KG results with RGF results, the latter being obtained using both standard and highly doped leads. Figures~\ref{eva} (a) and (b) compare the conductances far away from the CNP, showing that the doping of the leads does not significantly affect the conductance. However, closer 
to the CNP, as shown in Figs. \ref{eva} (c) and (d), the RGF results obtained with highly doped leads are much closer to the KG results. Thus the metallic leads provide a significant enhancement of the conductance of the complete system, even when the electronic states in the device are localized. This signifies the importance of the leads also when studying long and narrow systems. On the other hand, as seen from the slope of the conductance, the number of channels in the leads does not affect the localization length of the device.

\begin{figure}
\begin{center}
  \includegraphics[width=\columnwidth]{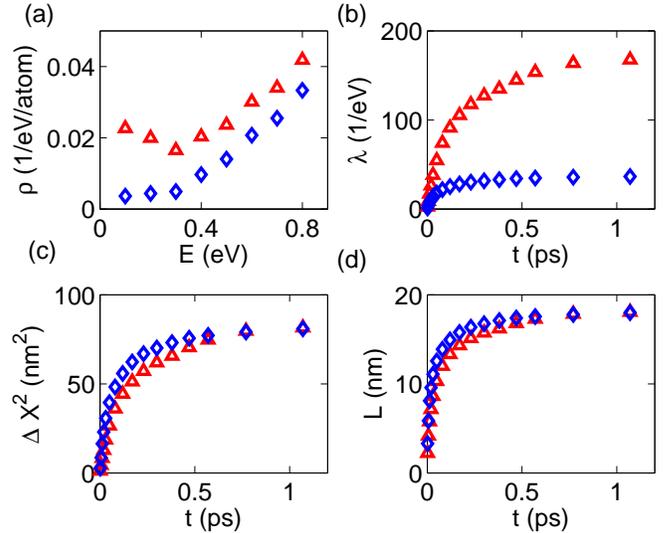}
  \caption{(Color online) Average (triangles) and typical (diamonds) values for various quantities for a 65-AGNR, as obtained using the KG method. (a) DOS as a function of energy, (b-d) $\lambda (E,t)$, MSD and propagating length at $E=0.2$ eV as a function of time. The defect concentration is $1 \%$ and the ensemble size is 10000.}
  \label{ratios}
\end{center}
\end{figure}

\section{Relation between the saturated value of the mean square displacement and the localization length}

As the MSD eventually converges to a finite value, one may ask whether this value is related to the localization length? It has been proposed that the square root of the converged MSD could be used as a definition of $\xi$ \cite{Triozon2000, Takada2013}. However, it has not been shown how this value is related to the localization length obtained from the exponential decay of the conductance. To answer the question, we first investigate the statistical properties of the MSD. It is known that the DOS tends toward a log-normal distribution in the localized regime.\cite{Schubert2009} This is illustrated in Fig.~\ref{ratios} (a), which shows the average and typical values of $\rho(E)$ for a 65-AGNR having a defect concentration of $1 \%$, with the typical values being significantly smaller than the average ones. A similar difference is seen between the typical and average values of $\lambda(E,t)$, as shown in Fig. \ref{ratios} (b). The MSD, which equals the ratio of $\lambda(E,t)$ and $\rho(E)$, may thus follow some other 
distribution than these quantities. It turns out that according to our simulations, in the localized regime, the average and typical values of the MSD coincide with good accuracy, as shown in Fig. \ref{ratios} (c). Thus the saturated value of the MSD may be obtained directly using Eq. (\ref{MSD}), which utilizes the trace operation. This operation can be approximated by using a few random vectors \cite{Weisse2006}, which is essential for achieving linear-scaling and thus computationally more efficient than studying a large ensemble of defect configurations. To further justify the equality, we have plotted the average and typical values of $L(E,t)$, given by Eq. (\ref{length}), in Fig. \ref{ratios} (d). As the propagating length is proportional to the square root of the MSD, the agreement between the values is even better.

\begin{figure}
\begin{center}
  \includegraphics[width=0.48\columnwidth]{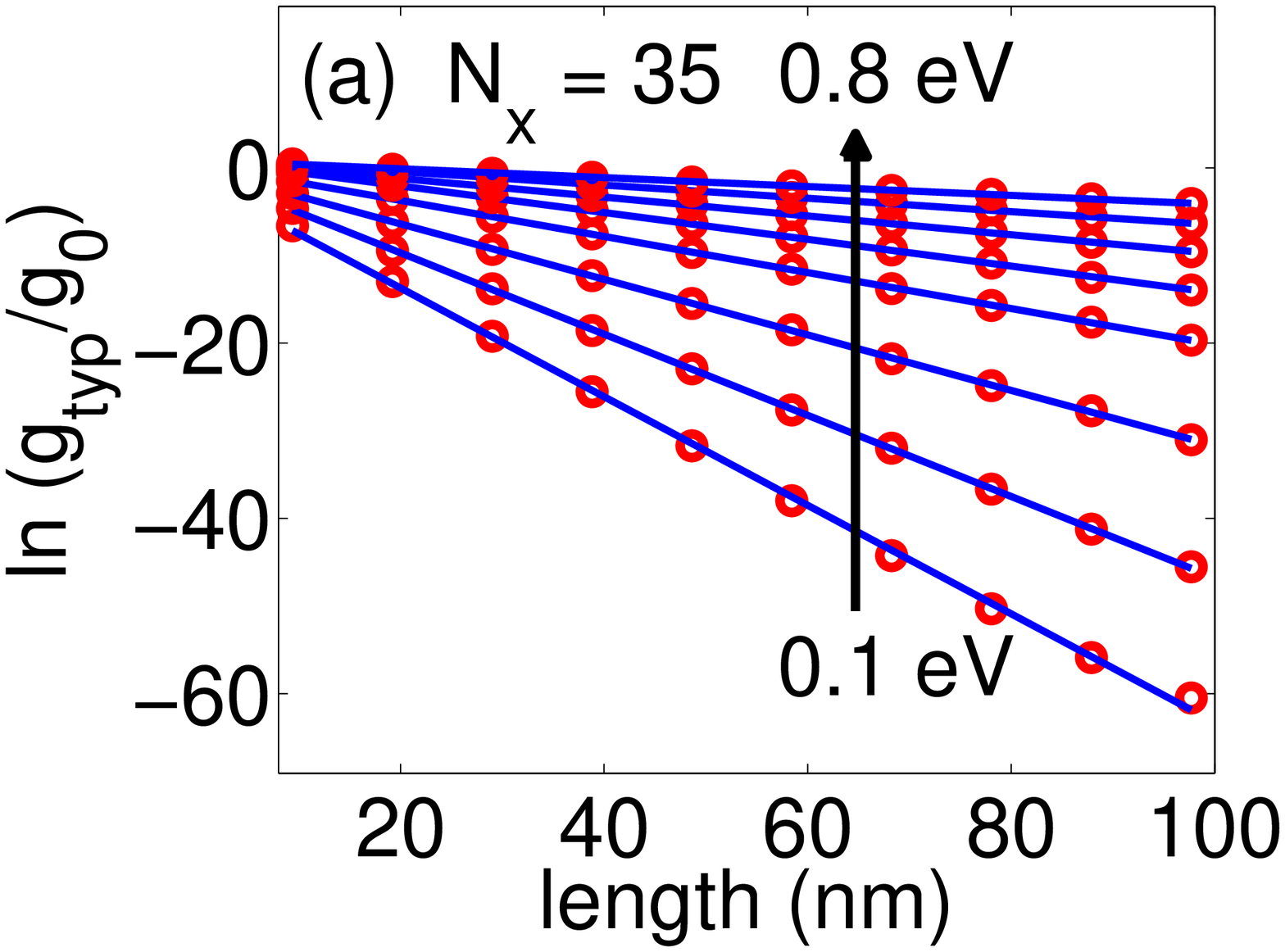}
  \includegraphics[width=0.48\columnwidth]{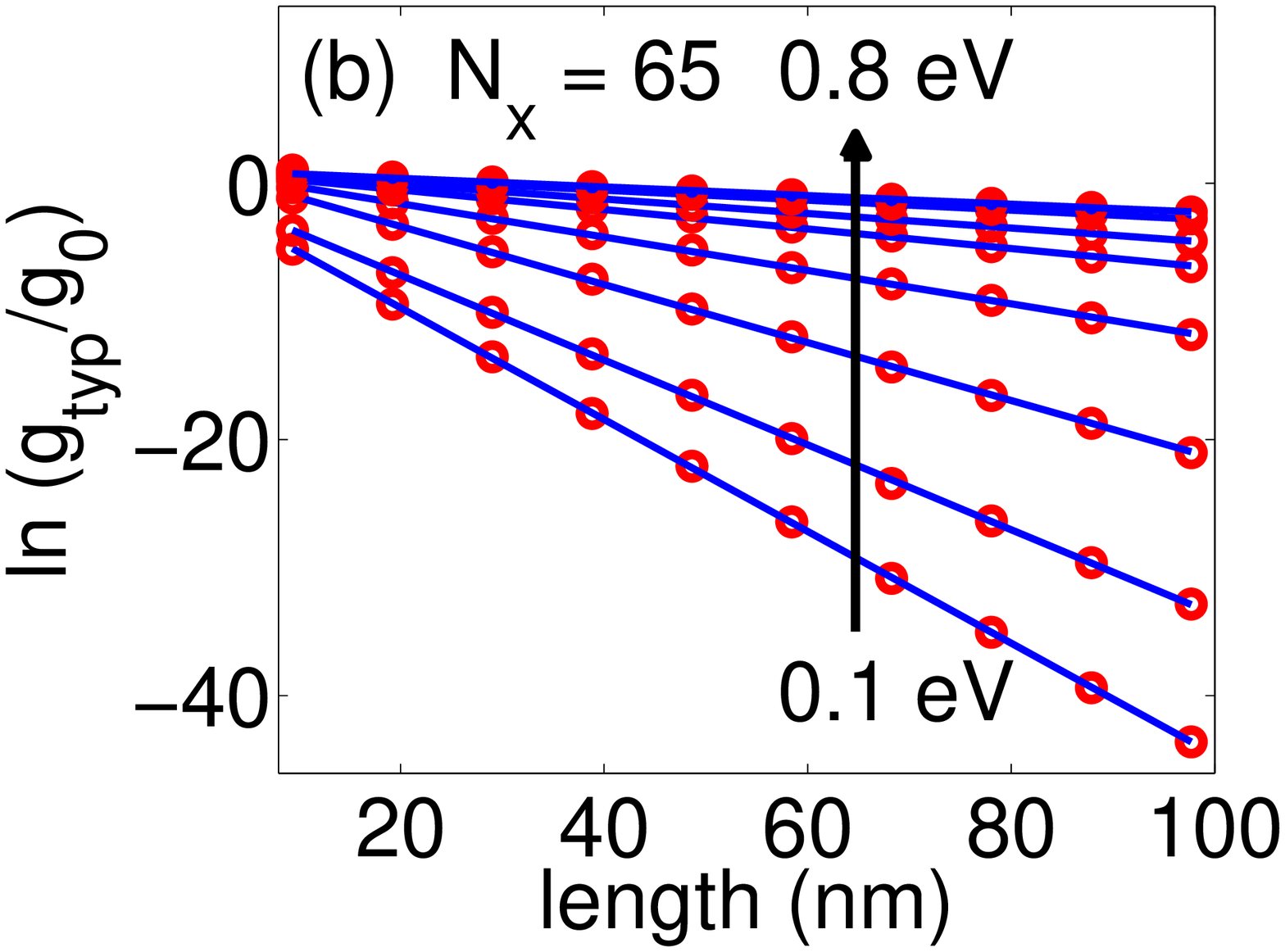} \\
  \includegraphics[width=0.48\columnwidth]{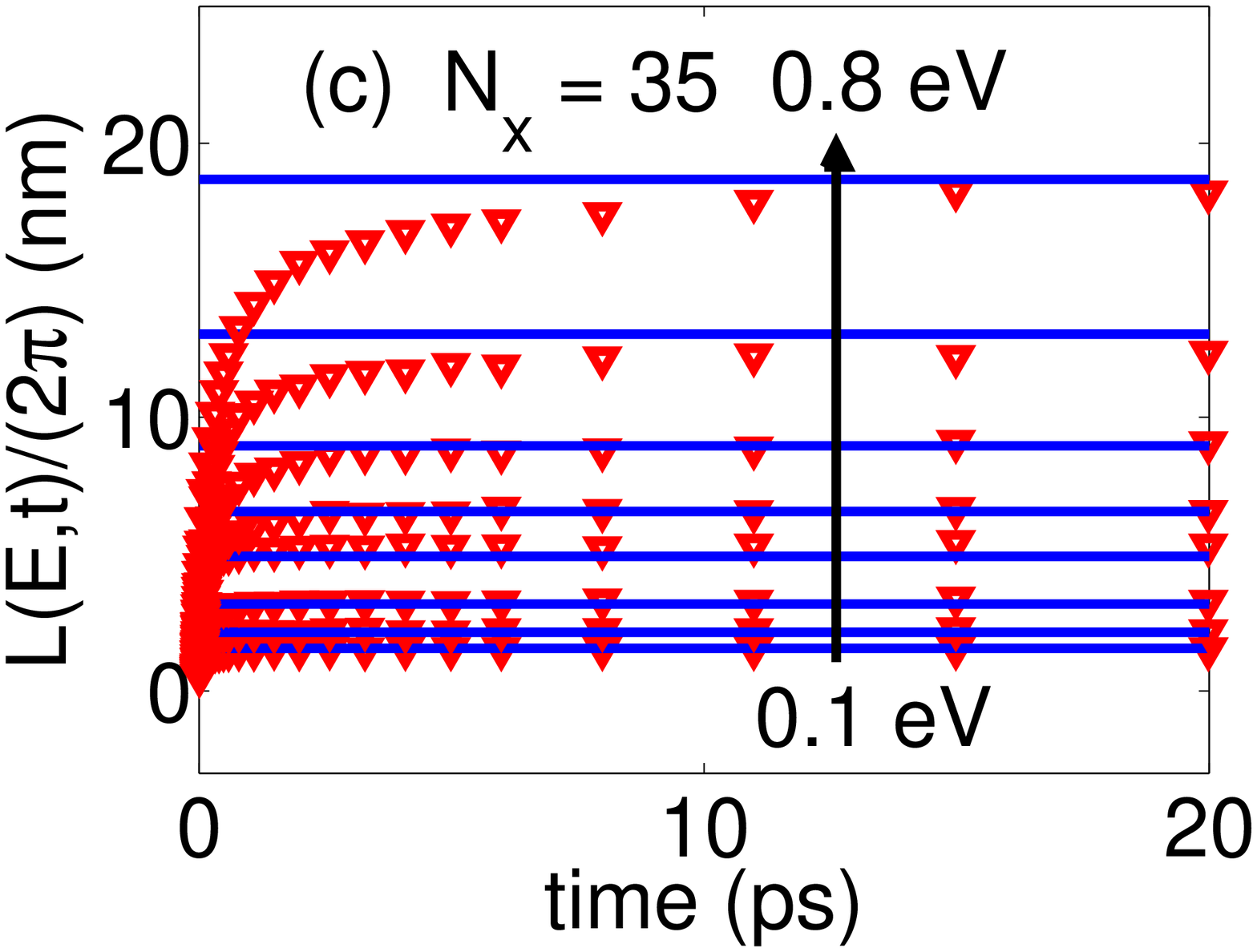}
  \includegraphics[width=0.48\columnwidth]{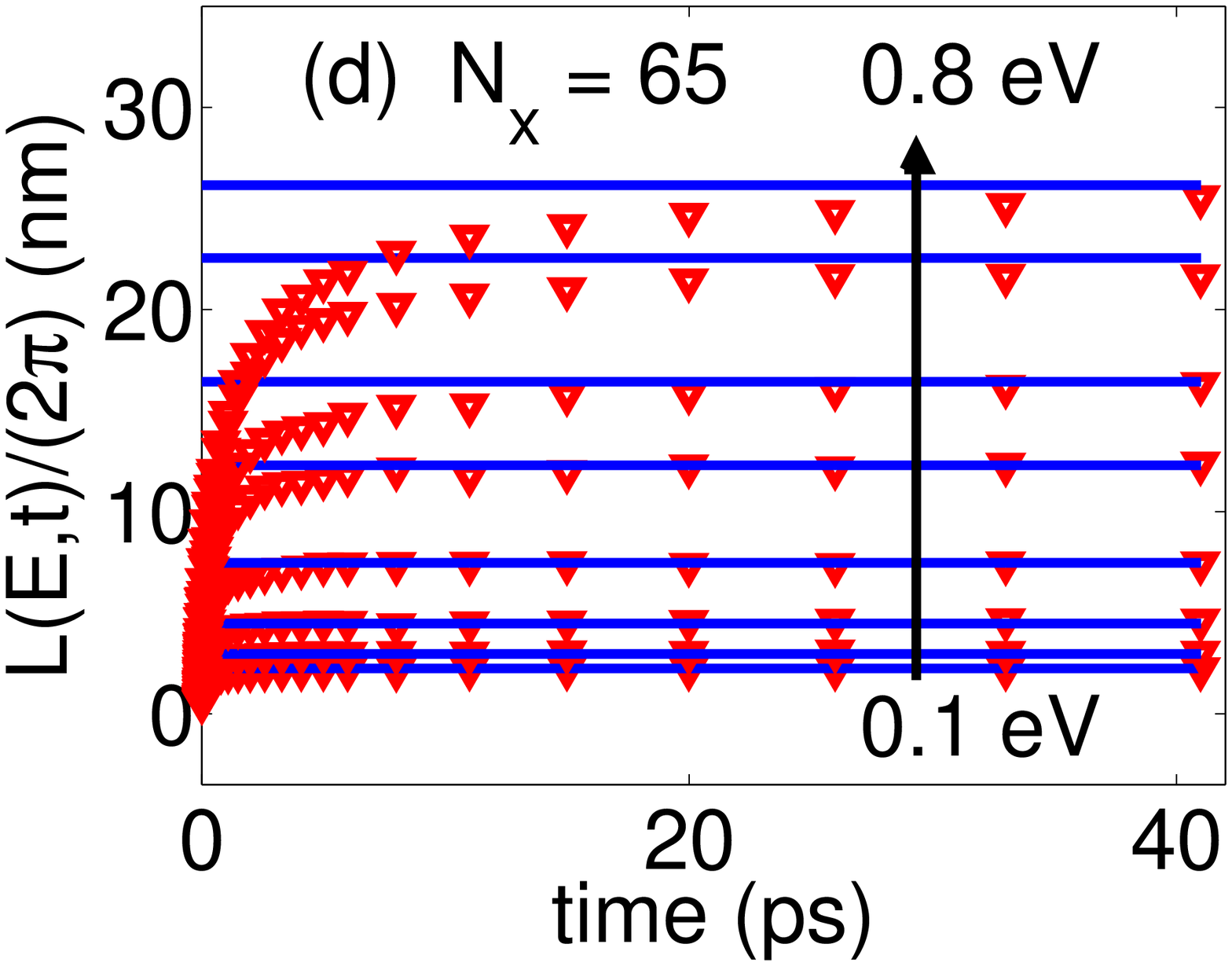} \\
  \includegraphics[width=0.48\columnwidth]{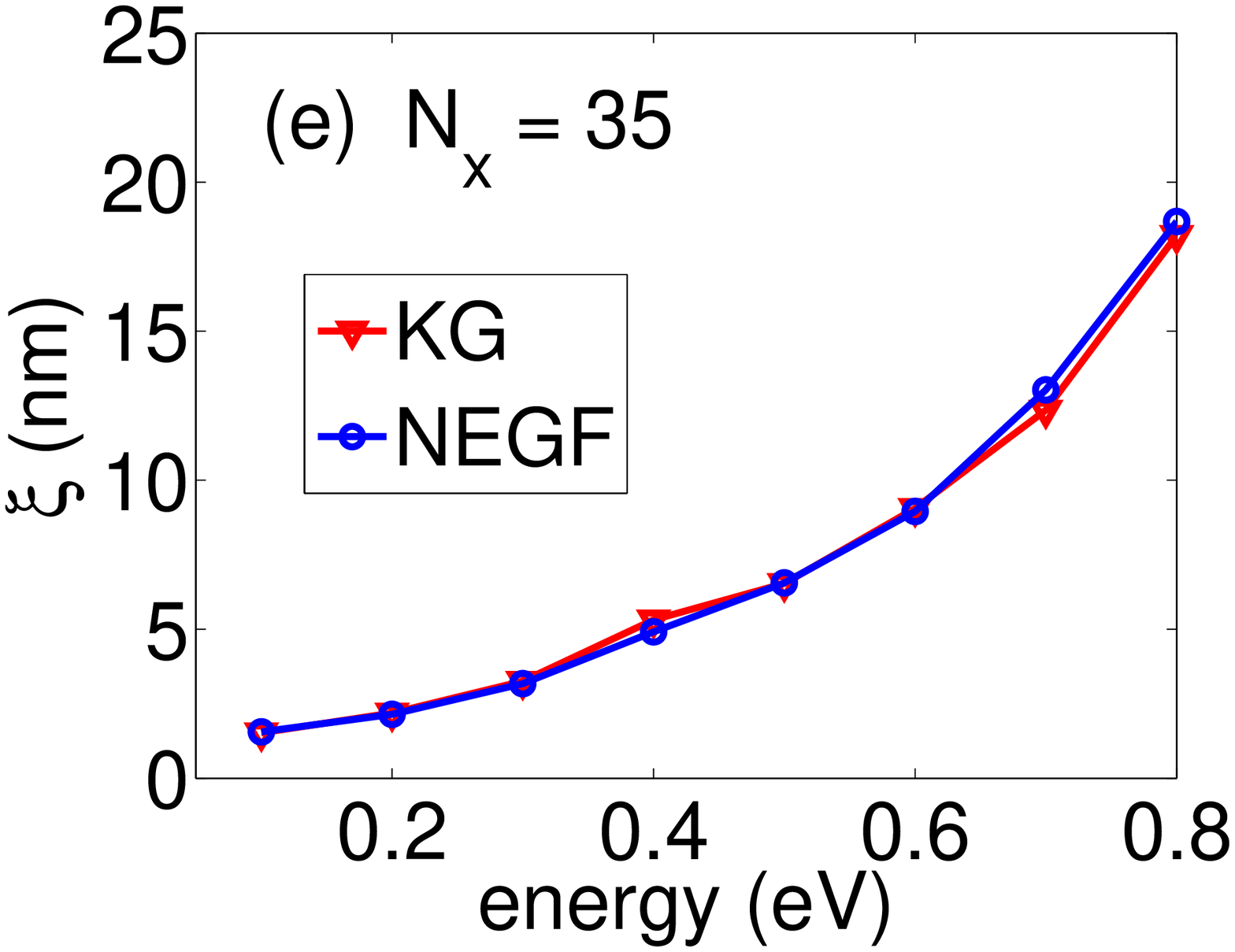}
  \includegraphics[width=0.48\columnwidth]{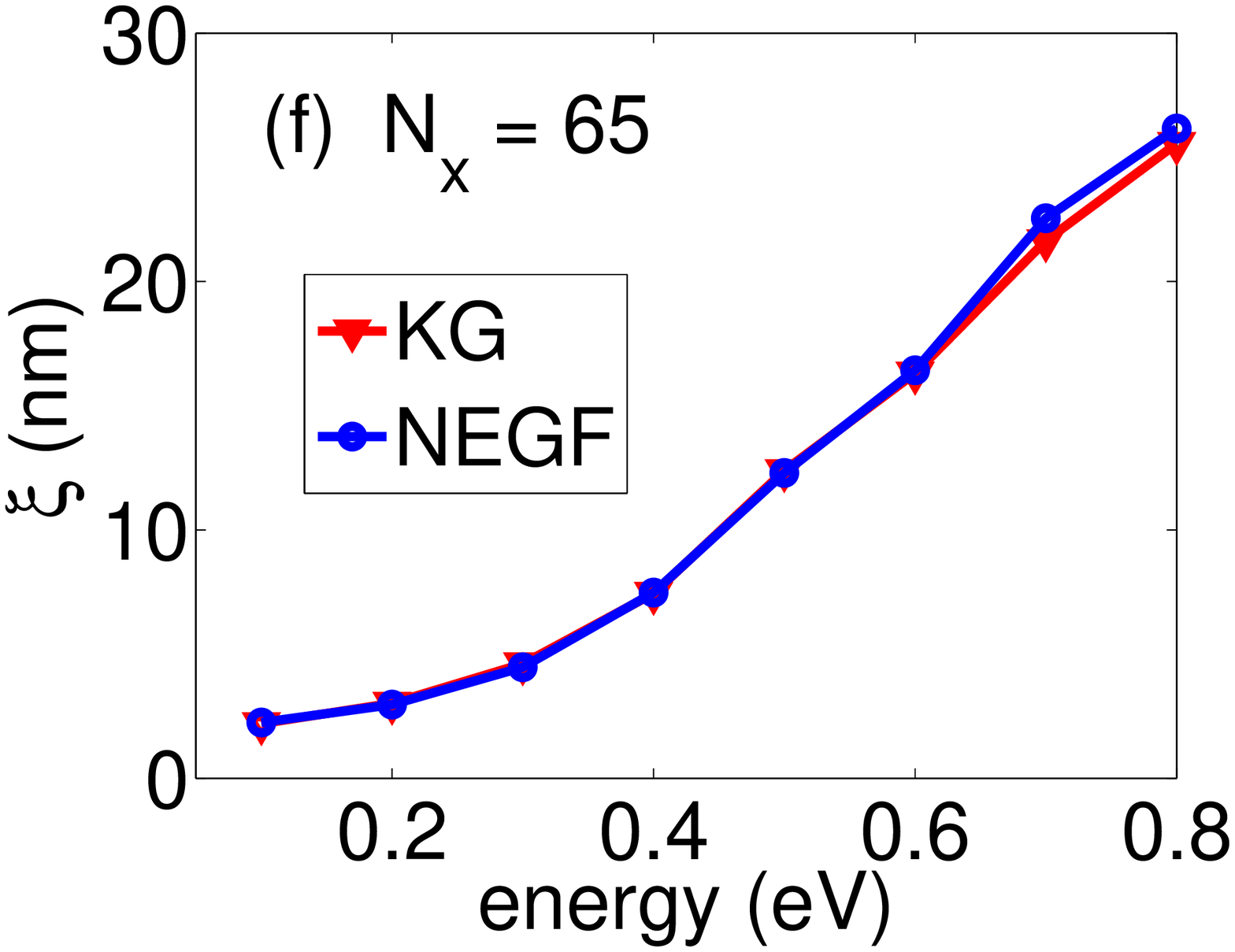}
  \caption{(Color online) (a) and (b): Linear fits to the typical conductances obtained from RGF simulations for a 35-AGNR and a 65-AGNR, respectively. The ensemble size is 5000. (c) and (d): Comparison of the time evolution of $\sqrt{ \Delta X^2(E, t)}/\pi$ at various energy points with the localization lengths (solid bars) obtained using the RGF method. (e) and (f): Comparison of the saturated values as a function of energy, given by the KG method, with the localization lengths obtained from panels (a) and (b). The defect concentration is $1 \%$ in all systems. The arrows indicate increasing and equally separated energy values.}
  \label{msd_sat}
\end{center}
\end{figure}

\begin{figure}
\begin{center}
  \includegraphics[width=\columnwidth]{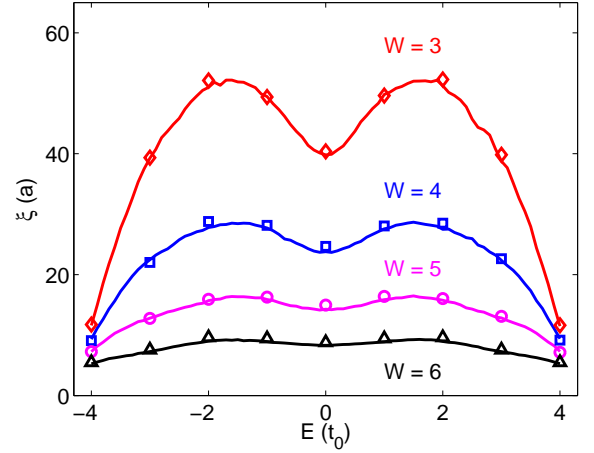}
  \caption{(Color online) Localization length for square lattice (with width $32a$, $a$ being the lattice constant) in the tube geometry (i.e., using periodic boundary conditions along the transverse direction) with Anderson disorder of different strengths. The solid lines have been obtained using the KG formalism and Eq.~\ref{xi_eq}, and the markers using exponential fits to RGF results, using an ensemble of 5000 different realizations of the disorder.}
  \label{sq_lattice}
\end{center}
\end{figure}

Being equipped with an efficient method to obtain the MSD, we are able to perform an extensive comparison of the saturated values of the propagating lengths with the localization lengths obtained from the RGF method. Figures \ref{msd_sat} (a) and (b) show exponential fits to the typical conductances of a 35-AGNR and a 65-AGNR, obtained by RGF simulations using ensembles of 5000 different defect configurations. Figures \ref{msd_sat} (c) and (d) demonstrate a comparison of the fitted values of the localization length obtained from (a) and (b) with $\alpha L(E, t)/2$, computed using the KG formalism. The parameter $\alpha$ has been set to $1/\pi$, which gives good correspondence with the results, as $L(E, t)/2\pi$ converges toward the corresponding localization length for each energy point. This suggests us to express the localization length as
\begin{equation}
\xi(E) = \frac{1}{\pi} \lim_{t \to \infty} \sqrt{ \Delta X^2(E, t)}.
\label{xi_eq}
\end{equation}
In other words, the localization length extracted from Eq.~(\ref{xi_eq}) is consistent with the typical two-terminal conductance decaying as $\exp \left[ -L/\xi(E) \right]$. This is further demonstrated in Figs. \ref{msd_sat} (e) and (f), which show localization lengths for different energies obtained from Eq.~(\ref{xi_eq}) compared against RGF results. Based on our simulations, we cannot conclude whether $\alpha$ equals exactly $1/\pi$ or just a value close to $1/\pi$. 

To test the generality of Eq.~(\ref{xi_eq}), we have used it to obtain the localization length of square lattices with Anderson disorder. Here, we consider a tube geometry by using periodic boundary conditions along the transverse direction. Anderson disorder is realized by setting the diagonal elements of the tight-binding Hamiltonian of the square lattice to values distributed randomly between $[-W/2,W/2]$, with $W$ describing the strength of the disorder. Figure~\ref{sq_lattice} shows a comparison between the KG and the RGF results for different values of $W$, indicating practically equal results from both methods. We note that, apart from a factor of 2 resulting from a different definition of the localization length in terms of the typical transmission, our results for the square lattice are consistent with those obtained by MacKinnon and Kramer. \cite{MacKinnon1983}

\begin{figure}[t!]
\begin{center}
  \includegraphics[width=\columnwidth]{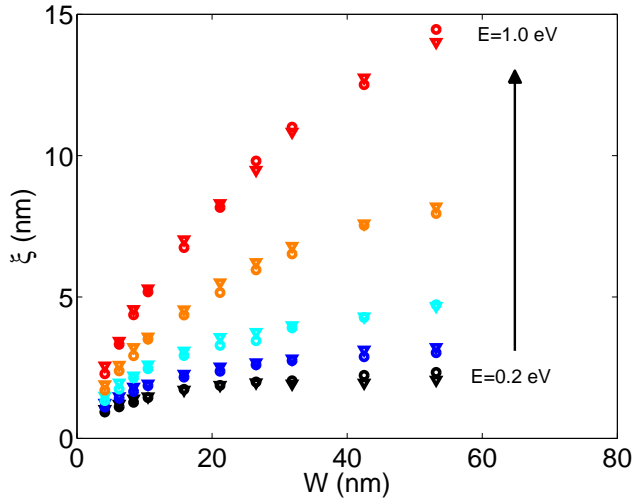}
  \caption{(Color online) Comparison of the localization lengths obtained using Eq. (\ref{xi_eq}) (triangles) and the RGF method (circles) as a function of system width, for ZGNRs with a defect concentration of $5 \%$. The ensemble size used in the RGF method is 5000. The arrow indicates increasing and equally separated energy values.}
  \label{1d2d}
\end{center}
\end{figure}

With the selected defect concentration of 1 \%, both the 35-AGNR and the 65-AGNR are quasi-1D systems, in the sense that $W$ is smaller than or of the same order as $\xi(E)$. To verify the generality of Eq.~(\ref{xi_eq}), we perform simulations in effectively two-dimensional systems, with $W \gg \xi$ and having a different edge termination. We compare the results given by Eq.~(\ref{xi_eq}) with RGF results for ZGNRs with a defect concentration of $5\%$ and with $N_y$ ranging form 14 to 250 and the corresponding $W$ from 3 nm to 53 nm. The results are shown in Fig. \ref{1d2d}, where it can be seen that when the width of the ribbon becomes larger than $\xi$, the scaling of $\xi$ starts to deviate from the quasi-1D behavior, which according to the Thouless relation\cite{Thouless1973, Avriller2006, Uppstu2012} is nearly proportional to $W$. As $W$ increases, the calculated values of $\xi$ approach limiting values, which may be interpreted as the corresponding localization lengths of a two-dimensional system. Thus our simulations indicate that Eq.~(\ref{xi_eq}) gives consistent results with the RGF method in both the quasi-1D and the effectively 2D regimes.

\section{Conclusions}

We have shown that localization properties of mesoscopic systems can be studied very efficiently using the Kubo-Greenwood method, by presenting direct comparisons with the recursive Green's function approach. Specifically, the results agree in both the one-dimensional and effectively two-dimensional regimes, with the width being much larger than the localization length in the latter case. We have shown that the typical and average values of conductance given by the Kubo-Greenwood and recursive Green's function methods agree up to a point where the propagating length defined in terms of the mean square displacement in the former method roughly equals four times the localization length, which allows the use of direct fitting to obtain the localization length. This fitting is not completely unambiguous, however, due to the eventual saturation of the propagating length. The convergence provides a more efficient method to compute the localization length, however, as the saturated value of the 
propagating length is directly proportional to it. When using this method to obtain the localization length, there is no need to distinguish typical and average values for the propagating length, which are found to be roughly equivalent to each other, and random vectors can be used instead of single-site wave functions to achieve linear-scaling in the Kubo-Greenwood method.  We have also discussed differences in the conductance given by the Kubo-Greenwood method and the recursive Green's function method close to the charge neutrality point, showing that with highly conducting leads, the recursive Green's function method gives results that are close to those acquired using the Kubo-Greenwood formalism.

\begin{acknowledgments}
This research has been supported by the Academy of Finland through its Centres of Excellence Program (project no. 251748).
\end{acknowledgments}
%

\end{document}